\documentstyle[aps,multicol,epsf]{revtex}

\begin{document}
\draft
\title{ Bak--Sneppen model near zero dimension }

\author{S.N. Dorogovtsev$^{1, 2, \ast}$, J.F.F. Mendes$^{1,\dagger}$, and Yu.G. Pogorelov$^{1,\ddagger}$}

\address{
$^{1}$ Departamento de F\'\i sica and Centro de F\'\i sica do Porto, Faculdade de Ci\^encias, 
Universidade do Porto\\
Rua do Campo Alegre 687, 4169-007 Porto -- Portugal\\
$^{2}$ A.F. Ioffe Physico-Technical Institute, 194021 St. Petersburg, Russia 
}

\maketitle

\begin{abstract}
We consider the Bak--Sneppen model near zero dimension where the avalanche exponent $\tau$ is close 
to $1$ and the exponents $\mu$ and $\sigma$ are close to $0$. We demonstrate that 
$\tau-1=\mu-\sigma=\exp\{-\mu^{-1}-\gamma+\ldots\}$ in this limit, where $\gamma$ is the Euler's constant. The avalanche hierarchy equation is rewritten in a form that makes possible to find the relation between the critical exponents $\sigma$ and $\mu$ with high 
accuracy. We obtain new, more precise, values of the critical exponents for the 1D and 2D Bak--Sneppen model and for the 1D anisotropic Bak--Sneppen model.   
\end{abstract}


\begin{multicols}{2}

\narrowtext


Perhaps the most simply formulated model showing avalanche behavior is the Bak--Sneppen model \cite{bs,pbak,pacz,masl95,fsb93,bdfjw94,mars94,bjw95}: "What could be simpler than replacing some random numbers with some other random numbers?" \cite{pbak}. Nevertheless, the exact solution of the Bak--Sneppen model is unknown even in one dimension. The value of the most fundamental quantity, i.e. of the upper critical dimension is also still under discussion \cite{rios98,boet}.

The formulation of the model is short indeed. 
A random number $f_i$ from some distribution ${\cal P}(f)$ is placed at an each site $i$ of a lattice. 
One replaces simultaneously the smallest number $f_{min}$ of them and the 
random numbers at its nearest neighbour sites by new random 
numbers from the distribution ${\cal P}(f)$, and afterwards the process is repeated. 

Avalanches in the Bak--Sneppen model are defined in the following way.  
The $f$-avalanche is defined as the sequence of the steps at which $f_{min}$ 
remains smaller than the given parameter $f$. (One may find a more detailed definition in \cite{pacz}.)

A very significant step to the understanding of the nature of the avalanches in this model was made in \cite{masl96} by S. Maslov who introduced the so called avalanche hierarchy equation for the distribution $P(s,f)$ of $f$-avalanche sizes $s$ (i.e. of temporal durations). From this {\em exact} equation, one may get the additional relation between the critical exponents of the model. Unfortunately, the exact solution of the equation is known only for the mean field situation. 
Two first terms of the expansion from the mean field solution, i.e. from the upper critical dimension, were calculated in \cite{mars98}. The precision of the results obtained by direct numerical integration of the avalanche hierarchy equation in its original form \cite{masl96} is only comparable with the precision of the Monte Carlo simulations \cite{pacz,gras}. 

There is another way to get analytical results.
It seems natural to start from the lower critical dimension which equals zero for the Bak--Sneppen model to make something similar to the well known $2+\epsilon$ expansion. Here, 
we derive from the avalanche hierarchy equation some convenient relations which enable us find the singular relation between the critical exponents near zero dimension and get values of the exponents at integer dimensions. Traditionally, one relates the exponents $\tau$ and $\mu$ \cite{mars98,rios98} (see the definition of these exponents below). 
The total curve $\tau(\mu)$ with the particular points for the integer dimensions and the areas of applicability of the approaches of \cite{mars98} and ours is depicted in Fig. 1.

For the distribution ${\cal P}(f)=e^{-f}, f>0$ \cite{dist}, the avalanche hierarchy equation
is of the form \cite{masl96}

\begin{equation}
\label{e1}
\frac{\partial P(s,f)}{\partial f} = \sum_{t=1}^{s-1} t^\mu P(t,f)P(s-t,f) - s^\mu P(s,f) \, .
\end{equation}

Here, in the scaling region, $s^\mu$ gives the average number of the distinct sites updating during an avalanche of the size $s$, where $\mu=d/D_f$, $d$ is the dimension of the lattice, $D_f$ is an avalanche fractal dimension \cite{pacz}. 
The physical meaning of the equation describing the hierarchical nature of avalanches in the Bak--Sneppen model is the following. The distribution $P(s,f)$ changes while $f$ grows because of two reasons. First, two consecutive avalanches of size $t$ and $s-t$ contribute to the avalanche of size $s$ (the second avalanche starts from one of the sites changed during the first avalanche that gives the factor $t^\mu$ in the sum). Second, some avalanches of size $s$ merge into a larger avalanche.

For the problem under consideration, the exponent $\mu$, which equals $0$ at the 
lower critical dimension and equals $1$ at the upper critical dimension 
(simulation in \cite{rios98} and \cite{boet}, gave different 
values for it, $d_{uc}= 8$ and $4$, correspondingly), 
is the given parameter. 
All other exponents are related with $\mu$ by Eq. (\ref{e1}). Below the threshold $f_c$ (i.e. in the symmetric phase) the solution of Eq. (\ref{e1}) has the following scaling form: 

\begin{equation}
\label{e2}
P(s,f)=s^{-\tau} F\left(s^\sigma (f_c - f) \right) \, .
\end{equation}

Eq. (\ref{e1}) resembles nonlinear differential equations with a peaking regime \cite{peak}. For such equations, it is possible to find both exponents included in Eq. (\ref{e2}). 

In \cite{mars98}, it was proposed to search for the Laplace transform of the 
 distribution $P(s,f)$:
\begin{equation}
\label{e3}
p(\alpha,f) = \sum_{s=1}^{\infty} P(s,f) e^{-\alpha s} \, .
\end{equation}

Then Eq. (\ref{e1}) gives
\begin{eqnarray}
\label{e4}
& & -\frac{1}{1-p(\alpha,f)}  \frac{\partial p(\alpha,f)}{\partial f} =
\sum_{s=1}^{\infty} P(s,f) s^\mu e^{-\alpha s} =
\nonumber 
\\[3ex]
& & (-1)^\mu \frac{\partial^\mu p(\alpha,f)}{\partial \alpha^\mu} = 
- \, \frac{1}{\Gamma(1-\mu)}\!\int_0^\infty \! dt \, t^{-\mu}\frac{\partial p(\alpha+t,f)}{\partial \alpha} ,
\end{eqnarray}
where $\partial^\mu/\partial\alpha^\mu$ denotes the fractional partial derivative ($\mu$ is certainly noninteger) and the last expression is its integral representation. $\Gamma(\ )$ is the gamma-function. The scaling relation for the solution of Eq. (\ref{e4}) is

\begin{equation}
\label{e5}
p(\alpha,f) = 1 - \alpha^{\tau-1} h \left( \frac{f_c - f}{\alpha^\sigma} \right) .
\end{equation}
Inserting Eq. (\ref{e5}) into Eq. (\ref{e4}) one obtains the usual relation between the critical exponents  
\begin{equation}
\label{e5a}
\tau=1+\mu-\sigma
\end{equation}
and the following integral-differential equation for the scaling function $h(x)$:

\begin{equation}
\label{e6}
\Gamma(1-\mu)x \frac{h^\prime(x)}{h(x)} = 
\int_0^x  dy\frac{y h^\prime(y)-\frac{\mu-\sigma}{\sigma}h(y)}{[1-(y/x)^{1/\sigma}]^\mu} \,  .
\end{equation}
Here $h^\prime(x) \equiv dh(x)/dx$.

In \cite{mars98}, Eq. (\ref{e6}) was used to obtain the expansion from the mean field solution \cite{fsb93,bdfjw94,bjw95} but it seems to be inconvenient. Let us show that one may transfer it to a purely integral form. 
The following lines demonstrate how that integration may be done.

\end{multicols}
\widetext
\noindent\rule{20.5pc}{0.1mm}\rule{0.1mm}{1.5mm}\hfill
\begin{eqnarray}
\label{e7}
& & \Gamma(1-\mu)  \frac{d\ln h(x)}{d x}  
\nonumber 
\\[3ex]
& & = \int_0^1  \frac{dz}{(1-z^{1/\sigma})^\mu} \left[xz \frac{dh(xz)}{d(xz)}-\frac{\mu-\sigma}{\sigma}h(xz)\right]  
=  \int_0^1  \frac{dz}{(1-z^{1/\sigma})^\mu} (xz)^{(\mu-\sigma)/\sigma+1}
\frac{d}{d(xz)}\left[(xz)^{-(\mu-\sigma)/\sigma} h(xz) \right] 
\nonumber 
\\[3ex]
& & = x^{(\mu-\sigma)/\sigma+1} \int_0^1  \frac{dz}{(1-z^{1/\sigma})^\mu}  
 z^{(\mu-\sigma)/\sigma+1} \frac{1}{z}\frac{d}{dx} \left[x^{-(\mu-\sigma)/\sigma}z^{-(\mu-\sigma)/\sigma} h(xz) \right] 
\nonumber 
\\[3ex]
& & = x^{(\mu-\sigma)/\sigma+1} \frac{d}{dx} x^{-(\mu-\sigma)/\sigma} 
\int_0^1 dz \frac{h(xz)}{(1-z^{1/\sigma})^\mu}  \, .
\end{eqnarray}
Applying $\int^x_0 dx$ to the first and the last lines of Eq. (\ref{e7}) and then integrating by parts (one may chose $h(x=0)=1$ \cite{mars98}) we get 

\begin{equation}
\label{e8}
\Gamma(1-\mu)  \ln h(x)  
= \int_0^1  \frac{dz}{(1-z^{1/\sigma})^\mu} 
\left[x h(xz) - \left( \frac{\mu-\sigma}{\sigma}+1 \right) \int_0^x du h(uz) \right]  
\end{equation}
and finally we obtain the equation for the scaling function $h(x)$ in the most convenient form:

\begin{equation}
\label{e9}
h(x) = \exp\left\{ \frac{1}{\Gamma(1-\mu)} 
\int_0^x \frac{dy}{[1-(y/x)^{1/\sigma}]^\mu} \left[ h(y) - \frac{\mu}{\sigma}\frac{1}{y}\int_0^y dz h(z) \right]  \right\}
\end{equation}
(if one does not demand $h(0)=1$, $h(x)$ in the left parts of Eqs. (\ref{e8}) and (\ref{e9}) is $h(x)/h(0)$).

Asymptotic form of $h(x)$ for large $x$ follows from the expansion of Eq. (\ref{e3}) in small $\alpha$. Below the threshold, $h(x)$ has to be

\begin{equation}
\label{e10}
h(x) \cong x^{(\mu-\sigma)/\sigma-1/\sigma} (c_0 + c_1 x^{-1/\sigma} + c_2 x^{-2/\sigma} + \ldots)  \, .
\end{equation}
This particular asymptotic behavior fixes the solution of Eq. (\ref{e9}) and the value of $\sigma$ for any given $\mu$. 
Substituting Eq. (\ref{e10}) into Eq. (\ref{e6}), Eq. (\ref{e8}), or Eq. (\ref{e9}), one gets the sume rule

\begin{equation}
\label{e11}
\int_0^\infty dx h(x) = \frac{1-(\mu-\sigma)}{\mu} \,\Gamma(1-\mu).   
\end{equation}
Note that if $h(x,\mu,\sigma)$ is a solution of Eq. (\ref{e9}) then $c\, h(c\, x,\mu,\sigma)$ is also a solution for any constant $c$. 

Eqs. (\ref{e9}) and (\ref{e11}) are the set of equations that lead to the scaling function $h(x,\mu)$ and $\sigma(\mu)$. Instead of Eq. (\ref{e11}), one may use equally the condition on the value of the exponent of the asymptote 

\begin{equation}
\label{e11a}
 [xh^\prime(x)/h(x)](x\to\infty) = \frac{(\mu-\sigma)-1}{\sigma} \, .  
\end{equation}
Hence, the problem is reduced to the eigen value problem for the nonlinear equation \cite{same}.

Let us study the solution of the system for small $\mu$. The expansion of the solution of Eq. (\ref{e9}) in  $x$ looks as

\begin{eqnarray}
\label{e12}
& & \ln h(x) = 
\left(1 - \frac{\mu}{\sigma} \right) B(\sigma,1-\mu) \left(\frac{\sigma x}{\Gamma(1-\mu)} \right) +
\left(1 - \frac{\mu}{\sigma} \right) B(\sigma,1-\mu) \left(1 - \frac{1}{2}\,\frac{\mu}{\sigma} \right)
B(2\sigma,1-\mu) \left(\frac{\sigma x}{\Gamma(1-\mu)} \right)^2 +
\nonumber 
\\[3ex]
& & \left(1 - \frac{\mu}{\sigma} \right)B(\sigma,1-\mu)
\left[\frac{1}{2}\left(1 - \frac{\mu}{\sigma} \right) B(\sigma,1-\mu) + \left(1 - \frac{1}{2}\,\frac{\mu}{\sigma} \right) B(2\sigma,1-\mu)  \right] \left(1 - \frac{1}{3}\,\frac{\mu}{\sigma} \right) B(3\sigma,1-\mu)
\left(\frac{\sigma x}{\Gamma(1-\mu)} \right)^3 + \ldots 
\nonumber 
\\[3ex]
& & = -\frac{\mu-\sigma}{\sigma}  \sum_{n=1}^{\infty} \frac{1}{n}B(\sigma,1-\mu) \ldots 
B(n\sigma,1-\mu) \left(\frac{\sigma x}{\Gamma(1-\mu)} \right)^n + \ldots \, ,
\end{eqnarray}
\hfill\rule[-1.5mm]{0.1mm}{1.5mm}\rule{20.5pc}{0.1mm}
\begin{multicols}{2}
\narrowtext
where $B(\ ,\ )$ is the beta-function. 
We shall see that the quantity $(\mu-\sigma)/\sigma$ is the smallest parameter of the problem near the lower critical dimension. If one tends formally $\mu$ to $0$, the last line of Eq. (\ref{e12}) tends to 

\begin{equation}
\label{e13}
\ln h(x) = -\frac{\mu-\sigma}{\sigma}\sum_{n=1}^{\infty} \frac{\Gamma(1+n\mu)}{n\,n!} 
\left(\frac{x}{\Gamma(1-\mu)}\right)^n
\end{equation}
and afterwards to

\begin{equation}
\label{e14}
\ln h(x) = -\frac{\mu-\sigma}{\sigma}\, \sum_{n=1}^{\infty} \frac{1}{n\,n! } 
\left(\frac{x}{\Gamma(1-\mu)} \right)^n .
\end{equation}
Thus, for small $\mu$, the solution $h(x)$ behaves as the following. For low enough $x$, the solution very slowly decreases from the value $h(0)=1$ and in some crossover region, $x \sim 1/\mu$, it comes to the asymptotic power tail, Eq. (\ref{e10}). 
We have to stress that three last limit equalities may be justified only for small $x$ and that the omitted terms in Eq. (\ref{e12}) do contribute to the power-law tail. Nevertheless, one may try to estimate $\sigma(\mu)$ for small $\mu$ inserting Eqs. (\ref{e13}) and (\ref{e14}) into the sum rule (\ref{e11}). The solution of the first of the equations obtained in such a way is $\mu-\sigma=\exp\{-\mu^{-1}-2\gamma+O(\mu)\}$, where $\gamma = 0.5772\ldots$ is the Euler's constant, and the solution of the second one is of the form 

\begin{equation}
\label{e15}
\mu-\sigma=\tau-1=\exp\{-\mu^{-1}-\gamma+O(\mu)\} \, .
\end{equation}
Thus, the dependence is non analytical but even the second term of the expansion can not be defined by such estimation.

In fact we failed to obtain the value of the constant analytically. Nevertheless, Eqs. (\ref{e9}) and (\ref{e11}) are very convenient for numerics, since iterations of Eq. (\ref{e9}) converge. (One may start, for instance, from the functions (\ref{e13}) or (\ref{e14}).) We checked the validity of the relation (\ref{e15}) for small $\mu$. The value of the constant in Eq. (\ref{e15}) obtained in such a way is $0.5771(5)$, i. e. that is the Euler's constant indeed. 

Solving Eq. (\ref{e9}) with the constraint, Eq. (\ref{e11}) or Eq. (\ref{e11a}), and the initial condition, one may easily get $\sigma(\mu)$ and $\tau(\mu)$ for any given $\mu$. The values of the exponent $\mu$ are known from simulation at integer dimensions with much higher precision than the values of $\tau$ because of the better available statistics \cite{pacz,gras}. Therefore we can improve essentially the precision of the known value of $\tau$. For the 1D Bak--Sneppen model, we get $\tau=1.0637(5)+0.4(\mu-0.4114)$, 
where $\mu=0.4114(2)$ is the value obtained from the Monte-Carlo simulation \cite{gras}.  
For the 2D Bak--Sneppen model, we obtain $\tau=1.229(1)+0.77(\mu-0.685)$, where $\mu=0.685(5)$ is the value obtained in \cite{gras}. 
(The last relations may be used to obtain better values of $\tau$ when the more precise values of the exponent $\mu$ will be available.)
Now the precision of $\tau$ coincides with that one of $\mu$. Note that these values are below the values of $\tau$ previously obtained from the simulation, $\tau(1D)=1.073(3)$ and $\tau(2D)=1.245(10)$ \cite{gras} but are in accordance with the less precise values found in \cite{mars98} by direct numerical solution of the avalance hierarchy equation, Eq. (\ref{e1}). The value of the exponent $\tau$ of the 3D Bak--Sneppen model 
may be obtained from the expansion from the upper critical dimension \cite{mars98}. In Fig. 1, we show the curve $\tau(\mu)$ together with the points for the integer dimensions and the low-$\mu$ asymptote,  (\ref{e15}), and the expansion from the upper critical dimension \cite{mars98}. 

Of course, the relation (\ref{e15}) is valid only for $\mu \ll 1$. Nevertheless, let us compare the value of $\tau$ at $\mu=0.4114$ obtained from Eq. (\ref{e15}), $\tau=1.0494$, with the calculated above $\tau(1D)$. One may see that these values are in qualitative agreement.   

In the case of the 1D anisotropic Bak--Sneppen model (i. e. for the update of the extremal site and only one its, for instance, right neighbour) the exponents $\sigma$ and $\mu$ are coupled by the following additional relation: $\sigma+\mu=1$ \cite{masl98}. Hence, one can find all the exponents of the problem. 
We obtained from  Eq. (\ref{e9}) the following value, $\mu=0.5779(5)$. In \cite{masl98}, 
two different values for $\mu$ were obtained, $\mu=0.58$ from Eq. (\ref{e1}) and $\mu=0.588$ found in another way. 
 The Monte-Carlo simulation made in \cite{vdb96} and \cite{hr98} gave  
$\mu=0.60(1)$ and $\mu=0.59(3)$ correspondingly. Therefore, we had to check our result. For that we solved numerically Eq. (\ref{e6}) with the initial condition $h(0)=1$ and the constraint, Eq. (\ref{e11}) or Eq. (\ref{e11a}). The result is $\mu=0.5778(5)$. Thus, the value $\mu=0.578$ seems to be more reliable but the question is still open. 

In summary, we have demonstrated that the simple transformation of the avalanche hierarchy equation made it convenient for analysis and numerics. We have obtained the non trivial singular relation, $\tau-1=\mu-\sigma=\exp\{-\mu^{-1}-\gamma+\ldots\}$ with the Euler's constant $\gamma$, between the scaling exponents of the Bak--Sneppen model near zero dimension. Using the known from simulation values of the exponent $\mu$, we have found, in fact, all other exponents of the Bak--Sneppen model in 1D and 2D with the same high precision. We have got also the exponents of the anisotropic 1D Bak--Sneppen model.

Nevertheless, one should note that the main problem of obtaining of the last independent critical exponent of the Bak--Sneppen model in a regular way remains open. \\

SND thanks PRAXIS XXI (Portugal) for a research grant PRAXIS XXI/BCC/16418/98. JFFM was partially supported by the project PRAXIS/2/2.1/FIS/299/94 and YGP was partially supported by the project PRAXIS/2/2.1/FIS/302/94. We also thank M.C. Marques for reading the manuscript and A.V. Goltsev  and A.N. Samukhin for many useful discussions.\\
$^{\ast}$      Electronic address: sdorogov@fc.up.pt\\
$^{\dagger}$   Electronic address: jfmendes@fc.up.pt \\
$^{\ddagger}$  Electronic address: ypogorel@fc.up.pt

\begin{figure}
\epsfxsize=61mm
\epsffile{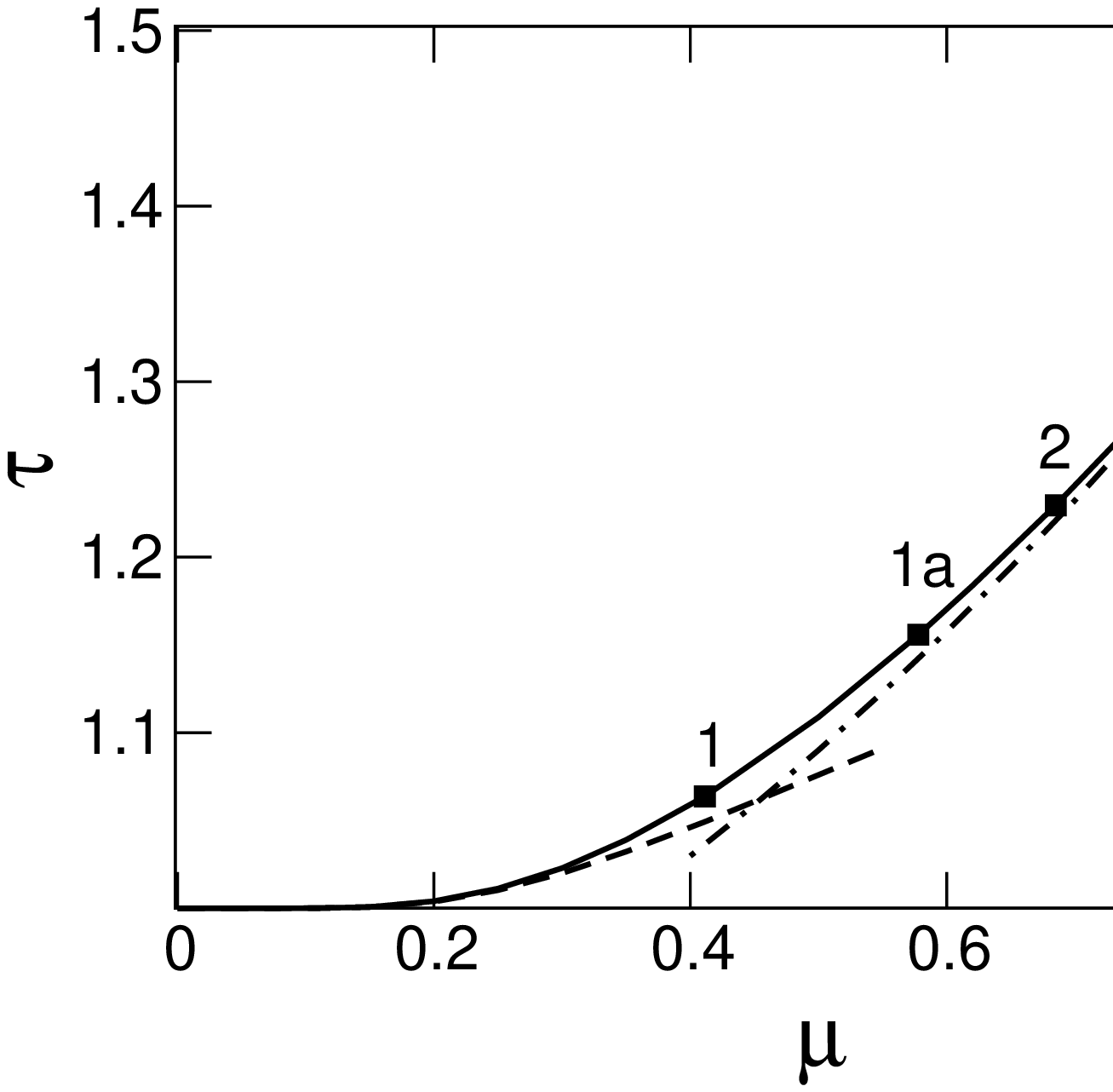}
\caption
{
The exponent $\tau$ vs. $\mu$ calculated from Eq. (\protect\ref{e9}) 
(see also \protect\cite{mars98,rios98}). 
The value of the 
exponent $\mu$ depends on the dimension $d$ of the system. $\mu(d=0)=0$, $\mu=1$ at 
the upper critical dimension.
The dashed line is obtained from Eq. (\protect\ref{e15}), i. e. by the expansion from 
the lower critical dimension, the dash-dotted line is 
the expansion 
\protect\cite{mars98} from the upper critical dimension. The points $1,2,$   
and $3$ correspond to the Bak--Sneppen model at 1D, 2D, and 3D. The point 
$1a$ corresponds to the 1D anisotropic Bak--Sneppen model.   
}
\end{figure}

\end{multicols}


\begin{references}
\bibitem{dist} Only the threshold value $f_c$, which is not interesting for us, depends on the particular form of the distribution ${\cal P}(f)$.
\bibitem{same} One may show that the result, $\sigma(\mu)$, will be the same if we omit the factor 
$\Gamma(1-\mu)$ in Eqs. (\protect\ref{e9}) and (\protect\ref{e11}).
\end{references}
\end{document}